\documentclass[rmp,letterpaper,twocolumn,noshowpacs,preprintnumbers,amsmath,amssymb,showkeys]{revtex4}

\usepackage{epsfig}
\usepackage{graphicx}
\usepackage{dcolumn}
\usepackage{bm}
\usepackage{bbm}

\begin{document}


\title{Physics and Derivatives: Effective-Potential Path-Integral Approximations of Arrow-Debreu Densities}

\author{Luca Capriotti $^{1,2}$} \email{luca.capriotti@nyu.edu} 
\author{Ruggero Vaia $^{3,4}$} \email{ruggero.vaia@isc.cnr.it} 

\affiliation{%
$^1$New York University, Tandon School of Engineering, 6 MetroTech Center, Brooklyn, NY 11201, United States of America \\
$^2$Department of Mathematics, University College London,  Gower Street, London WC1E 6BT, United Kingdom \\
$^3$Istituto dei Sistemi Complessi, Consiglio Nazionale delle Ricerche, via Madonna del Piano 10, I-50019 Sesto Fiorentino (FI), Italy\\
$^4$Istituto Nazionale di Fisica Nucleare, Sezione di Firenze,
             via G.~Sansone 1, I-50019 Sesto Fiorentino (FI), Italy.
}%

\date{\today}

\begin{abstract}
We show how effective-potential path-integrals methods, stemming on a simple and nice idea originally due to Feynman and successfully employed in Physics for a 
variety of quantum thermodynamics applications, can be used to develop an accurate and easy-to-compute semi-analytical approximation of transition probabilities and Arrow-Debreu densities for arbitrary diffusions. We illustrate the accuracy of the method by presenting results for the Black-Karasinski and the GARCH linear models,  for which the proposed approximation provides remarkably accurate results, even in regimes of high volatility, and for multi-year time horizons. The accuracy and the computational efficiency  of the proposed approximation makes it a viable alternative to fully numerical schemes for a variety of derivatives pricing applications.

\end{abstract}


\maketitle

\section*{Introduction}

Path integrals \cite{feynman2010quantum}, also known  as Wiener integrals in stochastic calculus \cite{wiener253, wiener294, kac1966}, are a well-established mathematical formalism
which has been used  for a long time in Physics to develop accurate approximations and  efficient computational techniques  \cite{kleinert2009path}. 

Among these, so-called semi-classical methods \cite{kleinert2009path}
play a central role. 
These  approximations can be developed in several ways which, while sharing the same limiting behavior, lead to genuinely different results.  
The renowned Wentzel-Kramers-Brillouin approximation \cite{Wentzel1926, Kramers1926, Brillouin1926}, which is equivalento to a saddle-point approximation of the path integral \cite{kleinert2009path, Rajaraman1975, Kakushadze2015}, and the Wigner-Kirkwood expansion \cite{Wigner1932,Kirkwood1933,FujiwaraOW1982,HilleryCSW1984}, 
are well-known theoretical devices in this context.

A prominent role among semi-classical approximations is played by so-called {\em effective potential} methods \cite{feynman2010quantum,feynman1998statistical} based, borrowing  renormalization group ideas, on `integrating out' the fluctuations around a `classical' trajectory.  Although exact in principle, the calculation can be performed only at some level of approximation, using a perturbation scheme in which the choice of the unperturbed system plays a crucial role in the quality of the approximation. 

A particularly successful effective potential approximation is the one stemming on a simple and nice idea originally due to Feynman \cite{feynman2010quantum} and independently developed by Giachetti and Tognetti \cite{GiachettiTognetti1985} and Feynman and Kleinert \cite{FeynmanKleinert1986} (GTFK), which is based on a self-consistent (non-local) harmonic approximation of the effective potential in a sense that will become clear in the following sections.

Basically, the GTFK effective potential is employed within the usual classical formalism, but accounts for the quantum nature of a system through suitable renormalization parameters it contains; hence, the approximation does not immediately lead to final results, but reduces a quantum-mechanical problem to a classical one, to be treated by any known method. Physicists know that this amounts to an enormous simplification. 

The most appealing aspect is that the classical behavior is fully accounted for by the GTFK potential, so it opened the way to face challenging quantum systems whose classical analogues were known to be characterized by peculiar nonlinear excitations, {\em e.g.}, those dubbed solitons in 1D or vortices in 2D. The latter are the `engine' of a topological phase transition, for the study~\cite{KosterlitzT1973} of which Michael Kosterlitz and David Thouless (KT) earned the 2016 Nobel prize. By the GTFK method it has been possible to establish that some real magnetic compounds do show a KT transition.

Other quantum systems that were succesfully treated by (suitable generalizations of) the same method, are frustrated antiferromagnets, {\em e.g.}, the so-called two-dimensional (2D) $J_1$-$J_2$ model~\cite{CapriottiFRT2004}, and 2D Josephson-junction arrays, which can be artificially fabricated, also with the inclusion of resistors; in the latter case, the effective potential could be naturally extended to account for the related dissipative coupling with the environment~\cite{1997CRTVpre}.

The connection between the so-called euclidean path integrals \cite{feynman2010quantum, kleinert2009path}, namely those employed to describe the thermodynamics of quantum systems, and the formalism of derivatives pricing has also been known since the seminal papers of \cite{Linetsky1997} and \cite{BennatiRosaClot1999} (see also the recent review \cite{Kakushadze2015}).  In particular, it is a known fact that a variable following a non-linear diffusion process can be described by the same formalism used to model the finite-temperature properties of a quantum particle in a potential which is linked to the drift of the diffusion, where the role of the mass is played by the inverse of the volatility squared, that of the temperature by the inverse of time and that of quantum fluctuations  by the Brownian noise \cite{BennatiRosaClot1999}. The interest in financial engineering for the path-integral formalism mainly  stems from the possibility of developing accurate approximation schemes, that are not otherwise available, or known, in traditional formulations of stochastic calculus \cite{BennatiRosaClot1999, Kakushadze2015, Capriotti2006}.  

In this paper, we will consider the application of the GTFK method to generalized short-rate models of the form $r_t = r(Y_t)$ 
with $Y_t$ following the non-linear diffusion process specified by the following stochastic differential equation (SDE)
\begin{equation}\label{eq.diffusion}
dY_t =\mu_y(Y_t)\,dt + \sigma_y(Y_t) \,dW_t,
\end{equation}
for $t>0$, where $\mu(Y_t)$ and $\sigma(Y_t)$  are the drift and volatility functions, respectively, $Y_{0} = y_{0}$, and $W_t$ is a standard Brownian motion.  

Short-rate models are of paramount importance in financial modeling, providing the foundation of many approaches 
used for pricing of both interest rate and credit derivatives \citep{andersen2010interest,o2010modelling}. In particular, celebrated 
affine models \citep{duffie} like those of \citep{Vasicek1977},  \citep{hw} and \citep{cir}, play a prominent role. This is mainly due to their 
analytical tractability allowing one to derive closed-form expressions for fundamental building blocks like zero-coupon bonds 
or, in the context of default intensity models \citep{o2010modelling}, survival probabilities. 

Unfortunately, the availability of closed-form solutions comes often at the price of less than realistic properties of the underlying rates. For instance, Gaussian models
such as those of  \cite{Vasicek1977} and  \cite{hw}, when calibrated to financial data, typically imply that rates can assume negative values with sizable probabilities. While this can be possibly not a problem 
for interest-rate models, especially in a low interest-rate environment, it is not consistent with absence of arbitrage in the context of 
default intensity models \citep{o2010modelling}. On the other hand, square-root diffusions such as that of   \cite{cir} - while guaranteed to be non-negative - may give rise to distributions of the par swap rate, see \cite{andersen2010interest,MercurioGarch}, that do not admit values below a finite threshold and may be considered therefore unrealistic.

Unfortunately, more realistic models lacks the same degree of analytical tractability as that shown by affine models. As a result, although widely used in practice, 
their implementations rely on computationally intensive  partial differential equations  (PDE) or Monte Carlo (MC) methods for the calculation of  bond prices or survival probabilities. 
This is particularly onerous in the context of  multi factor problems, notably the ones involving the calculation of  valuation adjustments (XVA), cf. \cite{gregory}, that are currently very prominent in financial engineering.
Indeed, these applications require Monte Carlo simulations and, {\em e.g.}, the valuation of conditional bond prices or survival probabilities at different points 
of the simulated paths, which are expensive to compute for models that lack closed-form solutions for these quantities. In this context, 
reliable analytical approximations are particularly important to reduce the numerical burden associated with these computations.

More specifically, in this paper we will focus on developing approximations of the so-called (generalized) Arrow-Debreu (AD) densities, 
see \citep{andersen2010interest,karatzas1991brownian}, also known as Green's functions, which are the fundamental building 
blocks for pricing contingent claims. These are defined, in this setting, as 
\begin{equation} \label{eq.ad}
\psi^Y_\lambda(y_T, y_{0}, T) = \mathbb{E}\Big[\delta(Y_T-y_T)e^{-  \lambda \int_{0}^T du \,r_u } \Big|\, Y_{0} = y_{0} \Big],
\end{equation}
where $\lambda$ is a real number,  and $\delta(\cdot)$ is the standard Dirac's delta function.
This, for $\lambda =0$, gives the transition density, of paramount importance for maximum-likelihood estimations in econometrics \cite{sahalia1999}, such that 
\begin{equation}\label{trans}
\int_A dy_T \,\, \psi^Y_{0}(y_T, y_{0}, T) \equiv \mathbb{P}\left[Y_{T} \in A \,|\, Y_{0} = y_{0}\right]~.
\end{equation}

The price at time $t=0$ of a European option with expiry $T$ and payout of the form $P(r_T)$, 
\begin{equation}
V(0) = \mathbb{E}\Big[e^{-\int_{0}^T du \, r_u} P(r_T) \Big|\, Y_{0} = y_{0} \Big],
\end{equation}
can be obtained by integrating  the product of the payout function and the ($\lambda = 1$) AD density over all the possible values of the short rate 
at time $T$, namely
\begin{equation}\label{cc}
V(0) = \int dy_T\, \psi^Y_1(y_T, y_{0}, T) P(y_T) ,
\end{equation}
where the integration is performed over the range of the function $y_T=r^{-1}(r_T)$. In particular, the moment generating function for 
the random process $\int_{0}^T \-du\,r_u$  can be obtained for $P\equiv 1$,
\begin{equation}\label{eq.zeroad}
Z_\lambda(r_{0}, T) = \int dy_T \, \psi^Y_\lambda(y_T, y_{0}, T)~,
\end{equation}
which, for $\lambda =1$, gives the value at time $t=0$ of a zero-coupon bond with maturity $T$ \cite{andersen2010interest}.
In the context of default intensity models, where the default of a firm is modeled by the first arrival of a Poisson process with time-dependent intensity $r_t$,  
 \cite{o2010modelling}, 
Eq.~($\ref{eq.zeroad}$) for $\lambda = 1$ represents the survival probability up to time $T$, conditional on survival up to time $t=0$. This is the fundamental  building block for the evaluation of cash flows that are contingent on survival or default, see \cite{o2010modelling}.

The structure of the paper is as follows. We start by reviewing the formalism of the GTFK effective potential method in the context of the path-integral formulation of quantum statistical mechanics. We then make the connection between the formalism used in quantum Physics and the one used in finance by reviewing the path-integral formulation of AD densities for non-linear diffusion and we show how the GTFK approximation can be used in the mathematical setting of stochastic calculus in order to develop a semi-analytical approximation for the generalized AD densities (\ref{eq.ad}), and zero-coupon bonds (\ref{eq.zeroad}) for non-linear diffusion of the form (\ref{eq.diffusion}). Remarkably, the GTFK method, yielding exact results in the limit of zero volatility and time to maturity as any semi-classical approximation, is also exact whenever the drift potential is quadratic, which means it is exact, as we will recall, for the Vasicek \cite{Vasicek1977} and quadratic model \cite{Kakushadze2015}.  We finally illustrate the remarkable accuracy of the GTFK method for models for which an analytical solution is not available via the application to the so-called Black-Karasinski (BK) model \cite{BK} and the so-called GARCH linear stochastic differential equation (SDE)   \cite{MercurioGarch, EEGARCH}, both of particular relevance for the valuation of credit derivatives.

\section*{Effective Potential Approximation in Quantum Statistical Mechanics}

We start by recalling the path-integral formalism of quantum thermodynamics for a non-relativistic particle of mass $m$ described by the standard Hamiltonian
\begin{equation}
\hat{\cal H} = \frac{\hat p^2}{2m} +  V(\hat x)
\end{equation}
where $\hat x$ and $\hat p$ are the canonical coordinate and momentum operators such that $[\hat  x, \hat p] = i\hbar$, with $\hbar$ the reduced Planck's constant, 
and where $V(\hat x)$ is the potential the particle is subject to.

The quantum thermodynamical properties of the particle at temperature ${\cal T}$ can be described by the {\em density matrix} \cite{feynman2010quantum}, 
\begin{equation}\label{eq.pi0}
\hat \rho = e^{-\beta \hat{\cal H} }
\end{equation}
where $\beta = 1/ k_B {\cal T}$, with $k_B$ the Boltzmann's constant. The elements of the density matrix, in the coordinate representation, can be expressed
in terms of Feynman's path integral \cite{feynman2010quantum} as
\begin{align}
\rho(x_T,  x_0, T) &\equiv \langle x_T | \hat \rho | x_0  \rangle = \nonumber \\ & \int_{x(0) =x_0}^{x(T) = x_T} \hspace{-0.1cm}{\cal D} [x(t)] \, \,e^{S[x(t)]}~,
\label{eq.pi}
\end{align}
where the path integration is defined over all paths $x(t)$ such that $x(0) = x_0$ and $x(T) = x_T$, with $T = \beta \hbar$ the so-called {\em euclidean time} and 
the functional
\begin{equation}\label{eq.action}
S[x(t)] = - \frac{1}{\hbar} \int_{0}^T dt \left[ \frac{m}{ 2} \dot x^2(t) + V(x(t)) \right]~,
\end{equation}
is the {\em euclidean action}. The functional integration in Eq.~(\ref{eq.pi}), is formally defined as the limit for $N\to \infty$ of the expression 
\begin{align}
\left(\frac{m}{{2\pi \hbar \Delta t}}\right)^{N/2} \int \ldots \int \prod_{i=1}^{N-1}dx_i  \exp{\big[S(x_i,x_{i-1})\big]}~,
\end{align}
with $\Delta t = T / N$, $x_N\equiv x_T$  and 
\begin{align}
&S(x_i,x_{i-1}) = \nonumber \\  &-\frac{\Delta t}{\hbar }\left [ \frac{m}{2}\frac{(x_i - x_{i-1})^2}{\Delta t } + V((x_{i-1}+x_{i})/2)  \right]~.  
\end{align}

Although the evaluation of the path integral in Eq.~(\ref{eq.pi}) is possible just in a few cases for simple potentials, the formalism allows for new kinds of approximations. 
In particular, here we pursue an approximation stemming on an idea originally due to Feynman, that consists in classifying the paths according to an equivalence relation, and consequently decompose the integral into a first sum over all paths belonging to the same class, and a second one over all the equivalence classes. In particular, equivalent paths are those who share the average point, defined as the functional
\begin{equation}
\bar x[x(t)] = \frac{1}{T} \int_{0}^T \-\-dt \,\,x(t)~,
\end{equation}
so that each equivalence class is labelled by a real number $\bar x$ representing the common average point and we can factor out in Eq.~(\ref{eq.pi}) an ordinary integral over $\bar x$, namely
\begin{equation}\label{eq.PIADprice}
\rho(x_T, x_0, T) =  \int d\bar x  \,\, \rho_{\bar x}(x_T,  x_0, T)~, 
\end{equation}
where the {\em reduced density matrix}
\begin{align}\label{eq.reddens}
& \rho_{\bar x}(x_T,  x_0, T)   =  \nonumber \\ &  \int_{x(0) =x_0}^{x(T) = x_T} \hspace{-0.1cm}{\cal D} [x(t)] \delta \left(\bar x - \frac{1}{T} \int_{0}^T dt \,\,x(t) \right) \, \,e^{S[x(t)]}~,
\end{align}
represents the contribution to the path integral in  Eq.~(\ref{eq.pi})  that comes from those paths that have $\bar x$ as average point.

As the path integration has been reduced to paths belonging to the same class, we can develop a specialized approximation for each class. In particular, the GTFK method approximates the potential in the action Eq.~(\ref{eq.action}) with a quadratic potential in the 
displacement from the average point $\bar x$
\begin{equation}\label{eq.trialpot}
V_{\bar x}(x) = w(\bar x) + \frac{m}{2} \omega^2(\bar x) (x-\bar x)^2~,
\end{equation}
where the parameters $w(\bar x)$ and $\omega^2(\bar x)$ are to be optimized so that the {\em trial} reduced density matrix
\begin{align}\label{eq.trialreddens}
& \bar \rho_{\bar x}(x_T,  x_0, T)  \nonumber \\ &=  \int_{x(0) =x_0}^{x(T) = x} \hspace{-0.1cm}{\cal D} [x(t)] \delta \left(\bar x - \frac{1}{T} \int_{0}^T dt \,\,x(t) \right) \, \,e^{S_{\bar x}[x(t)]}~,
\end{align}
with the action given by
\begin{equation}\label{eq.trialaction}
S_{\bar x}[x(t)] = - \frac{1}{\hbar} \int_{0}^T dt \left[ \frac{m}{ 2} \dot x^2(t) + V_{\bar x}(x(t)) \right]~,
\end{equation}
best approximates the reduced density matrix in Eq.~(\ref{eq.reddens}). Note that one does not need to include a linear term in the trial potential~\eqref{eq.trialpot}, since it would give a vanishing contribution to the trial action~\eqref{eq.trialaction}, due to the very definition of $\bar{x}$.

The path integral in Eq.~(\ref{eq.reddens}), corresponding to the harmonic action (\ref{eq.trialaction}) can be worked out analytically \cite{PQSCHA},  giving
\begin{align}\label{eq.reddens2}
& \bar \rho_{\bar x}(x_T, x_0, T)  =  \sqrt{\frac{m}{2\pi\beta \hbar^2}} e^{- \beta w(\bar x)} \frac{f}{\sinh f} \times  \nonumber \\ 
&\frac{1}{\sqrt{2\pi\alpha}} \exp\left[ -\frac{\xi^2}{2\alpha}
 -\frac{m\omega\coth f}{4\hbar}\,(x_T-x_0)^2 \right]~,
\end{align}
where $\xi = (x_T+x_0)/2 - \bar x$, $f = \beta \hbar \omega(\bar x) /2$ and 
\begin{equation}\label{eq.alpha}
\alpha(\bar x) = \frac{\hbar}{2m\omega(\bar x)}\left (\coth f(\bar x) -\frac{1}{f(\bar x)} \right )~.
\end{equation}
The diagonal elements of the  reduced density matrix read in particular
\begin{align}\label{eq.reddens3}
& \bar \rho_{\bar x}(x_0,  x_0, T)  =  \sqrt{\frac{m}{2\pi\beta \hbar^2}} e^{-\beta w(\bar x)} \frac{f}{\sinh f} \times  \nonumber \\ 
&\frac{1}{\sqrt{2\pi\alpha}} \exp\left[ -\frac{\xi^2}{2\alpha} \right]~,
\end{align}
taking a suggestive form in terms of a Gaussian distribution with mean $\bar x$ and variance $\alpha(\bar x)$, describing the fluctuations around the average point.  
In particular, the so-called {\em partition function}, ${\cal Z}$, \cite{feynman1998statistical} assumes the classical form
\begin{align}\label{eq.part}
{\cal Z} \equiv & \int d\bar x \int d x_0 \,\, \rho_{\bar x}(x_0,  x_0, T)  = \nonumber \\
& \sqrt{\frac{m}{2\pi \beta \hbar^2}} \int d\bar x \,\,e^{-\beta \,V_{\rm eff}(\bar x)}~,
\end{align}
where the GTFK {\em effective potential} reads:
\begin{align}\label{eq.effpot}
V_{\rm eff}(\bar x) = w(\bar x) 
 + \frac{1}{\beta} \ln \frac{\sinh f(\bar x)}{f(\bar x)}~.
\end{align}

In order to close the approximation we still need to devise an optimization scheme for the parameters $w(\bar x)$ and  $\omega(\bar x)$ in Eq.~(\ref{eq.trialpot}).  For example, we could simply identify the trial potential (\ref{eq.trialpot}) with the expansion of $V(\bar x)$ up to second order 
by setting $w(\bar x) = V(\bar x)$ and $\omega(\bar x) = V^{\prime\prime}(\bar x)$ for any $\bar x$. However, this approximation has limitations. For instance, it can happen that  $V^{\prime\prime}(\bar x)$ is negative: in this case, writing  $f = \beta \hbar \omega /2$ as  $f= i\phi $, $\alpha$ can be analytically continued as $\alpha = (\beta \hbar^2 /4m)(1/\phi^2 - \cot \phi/\phi)$, which diverges to $+\infty$ for $\phi \to \pi^-$ (or $f^2 \to -\pi^2$) and is negative for $\phi > \pi$ ($f^2 < -\pi^2$). As a consequence, if $\omega^2(\bar x)$ is negative,
for sufficiently large time horizons $T$ we have $f^2 < -\pi^2$ and $\alpha(\bar x) < 0$. In this situation, the reduced density matrix  (\ref{eq.reddens2})  is not well defined and the approximation breaks down. 

A more robust approximation can be devised by observing that the Gaussian density $\bar\rho_{\bar x}(x_0,x_0,T)$ has to be close to $\rho_{\bar x}(x_0,x_0,T)$, so that $V_{\bar x}(x)$ must approximate $V(x)$ not only at $\bar x$: this is accomplished by requiring the equality of the Gaussian averages of the true and the trial potentials, and of their derivatives up to the second one 
\begin{align} 
\langle\!\langle V(\bar x +\xi) \rangle\!\rangle &=  \langle\!\langle V_{\bar x}(\bar x + \xi) \rangle\!\rangle \nonumber \\ & = w(\bar x) + \frac{m}{2}\omega^2(\bar x)\alpha(\bar x)~, \label{GTFK1}\\ 
\langle\!\langle V^{\prime\prime}(\bar x + \xi) \rangle\!\rangle&=\langle\!\langle  V_{\bar x}^{\prime\prime}(\bar x + \xi)\rangle\!\rangle = m\omega^2(\bar x)~,\label{GTFK2}
\end{align}
with the short-hand notation 
\begin {align}
\langle\!\langle F(\bar x + \xi) \rangle\!\rangle &\equiv \frac{1}{\sqrt{2 \pi\alpha(\bar x)}} \int_{-\infty}^{+\infty} d\xi \,\,e^{-\xi^2/2\alpha(\bar x)} F(\bar x + \xi) \nonumber \\
 &= e^{\frac{\alpha(\bar x)}{2}\partial_x^2} F(\bar x)~,
\end{align}
and $\alpha(\bar x)$ given by Eq.~(\ref{eq.alpha}).
The equations above impose that the expectation value according to the Gaussian probability distribution in Eq.~(\ref{eq.reddens3}) of the potential and of its second order expansion are in agreement with each other, for {\em every} value of $\bar x$. Under the GTFK approximation the quantum effects are embedded in the notion of the effective potential (\ref{eq.effpot}) which is a renormalized version of the potential $V(x)$ where $\alpha(\bar x) \equiv \langle\!\langle \xi^2 \rangle\!\rangle$ -- representing the average quadratic fluctuations around $\bar x$ due to the quantum effects -- is the renormalization parameter. Note that Eq.~\eqref{GTFK2} is self consistent, meaning that its solution $\omega^2(\bar x)$ in turn determines the variance~$\eqref{eq.alpha}$.

It can be shown that the above determination of the parameters $w({\bar{x}})$ and $\omega({\bar{x}})$ satisfies a variational principle based on the so-called Jensen-Feynman inequality, ${\cal Z}\ge {\cal Z}_0 \,e^{\langle S-S_0\rangle_0}$, where the functional average is taken with whatever trial action $S_0$, ${\cal{Z}}_0$ being the corresponding partition function. Indeed, taking $S_0=S_{\bar{x}}$ and maximizing the right-hand side of the inequality one just finds Eqs.~\eqref{GTFK1} and~\eqref{GTFK2}.

The GTFK method, becomes exact in both limits of high-temperature $\beta \to 0$ and vanishing quantum effects $\hbar/m \to 0$, for which the parameter $\alpha$ vanishes as $\beta\hbar^2/12m$ and the effective potential (\ref{eq.effpot}), coincides with the exact classical potential:
\begin{equation}\label{VeffhighT}
V_{\rm eff}(\bar x) =  V(\bar x) + \frac{\beta\hbar^2}{24m} V^{\prime\prime}(\bar x) + O(\beta^2\hbar^4/m^2),
\end{equation}
so that the partition function in Eq.~(\ref{eq.part}) coincides with the well-known exact classical result \cite{feynman1998statistical}.

The effective potential can be compared to the semiclassical effective potential introduced by Wigner and Kirkwood~\cite{Wigner1932,Kirkwood1933,FujiwaraOW1982,HilleryCSW1984} (WK), that was substantially found as an expansion in $\beta$ and $\hbar$ of the exact classical effective potential $V_{\rm ex}$, defined such that the quantum density bears the classical form
\begin{equation}
 \rho(x_0, x_0, T) \equiv \frac 1{\cal Z}~e^{-\beta V_{\rm ex}(x_0)} ~.
\end{equation}
The WK expansion is in principle exact, but only the first few terms are practically affordable, and while lowering the temperature all terms soon diverge. One has indeed~\cite{JizbaZ2014}
\begin{equation}
 V_{\rm ex}(x_0) = V(x_0)+\frac{\beta\hbar^2}{12m}V^{\prime\prime}(x_0)
   -\frac{\beta^2\hbar^2}{24m}{V^\prime}^2(x_0) +\dots
\end{equation}
This apparently disagrees from the expansion~\eqref{VeffhighT} because the comparison is a little subtle: indeed, $V_{\rm eff}$ has not to be directly compared with $V_{\rm{ex}}$, because, in order to obtain $\rho(x_0, x_0, T)$ one cannot integrate over $x_0$ as made in Eq.~\eqref{eq.part}, but rather over $\bar{x}$. Accounting for this~\cite{Kleinert1986}, the WK and the GTFK effective potentials do agree~\cite{1990VTijmpb,1992CTVVpra}.  Similarly, GTFK is distinct from the exponential power series expansion of \cite{makri}, previously applied successfully in the financial context \cite{Capriotti2006, EEBK, EEGARCH}, and which we will use as one of the benchmarks when discussing our numerical results.

With respect to these approaches, the GTFK method has a strong advantage: it still gives a meaningful representation of the thermodynamics down to zero temperature, where it is equivalent to the so-called self-consistent harmonic approximation~\cite{Koehler1966a,Koehler1966b}, that was initially applied to quantum crystal lattices. Therefore, increasing temperature from zero the accuracy increases more and more, because the renormalization parameter $\alpha(\bar x)$ decreases. The price to pay is that one still has to solve the classical problem with the effective potential, but this is nevertheless a huge simplification, especially in view of the plenty of methods that have been developed to treat classical systems. In particular, thanks to the fact that the nonlinear character of the potential is kept, the GTFK approach allows for studying quantum systems whose classical counterpart is characterized by nonlinear excitations (solitons, vortices) and constitutes a much simpler and clearly interpretable alternative to heavy numerical approaches, such as Quantum Monte Carlo.

The GTFK approach is also distinct from other semi-classical path-integral approximations, like the Wentzel-Kramers-Brillouin (WKB) \cite{Wentzel1926, Kramers1926, Brillouin1926} or the equivalent saddle-point approximations \cite{kleinert2009path, Rajaraman1975, Kakushadze2015}, which are based on a power-series expansion of the action around the classical trajectory $x_{\rm{c}}(t)$ rather than around the average point, {\em i.e.}, the density matrix, Eqs.~\eqref{eq.pi} and~\eqref{eq.action}, is expressed as
\begin{align}
 \rho(x_T,  x_0, T) = e^{S[x_{\rm{c}}(t)]} \int_{\tilde x(0) = 0}^{\tilde x(T) = 0}\!\!{\cal D} [\tilde x(t)] \, \,e^{\tilde S[\tilde x(t)]}~,
\label{eq.pi.wkb}
\end{align}
where $x_{\rm{c}}(t)$ obeys the classical equation of motion $\delta{S}/\delta{x(t)}=0$ and satisfies the boundary conditions $x_{\rm{c}}(0)=x_0$ and $x_{\rm{c}}(T)=x_T$, while the path summation is over closed paths $\tilde{x}(t)=x(t){-}x_{\rm{c}}(t)$ with the expanded action
\begin{equation}\label{eq.action.wkb}
 \tilde S[\tilde x(t)] = - \frac{1}{\hbar} 
 \int_{0}^T dt \left[ \frac{m}{ 2} \dot{\tilde x}^2(t) + \frac{V^{\prime\prime}(x_{\rm{c}}(t))}2\,{\tilde x}^2(t) +\dots\right]~.
\end{equation}
The WKB approximation is exact for a quadratic potential, and, the first term being of order $\hbar^{-1}$, it can include the effect of tunneling (for instance, in a double-well potential) at variance with the GTFK; however, one has to consider that it is not crucial to account for tunneling effects, as they are soon overwhelmed by quantum thermal fluctuations and are practically absent in many-body systems; moreover, beyond few relatively simple cases, the evaluation of the path integral~\eqref{eq.action.wkb} is generally hard, mainly due to the dependence of $\tilde{S}$ upon the classical path.
On the other hand, the non-local nature of the GTFK approximation yields the possibility of tuning two families of parameters, $w(\bar x)$ and $\omega(\bar x)$, allowing one to look for the best approximation of the true action in a richer space, while preserving the property of being exact in the classical limit and for harmonic actions. By `richer space' we mean that the trial action, thank to its dependence on the average-point functional, is much more general than the local actions corresponding to physical potentials. The GTFK can also be systematically improved, at least in principle, without suffering from the divergencies appearing instead in most perturbative approaches \cite{kleinert2009path}.

The generalizations of the GTFK approach to many degrees of freedom, as well as to Hamiltonian systems~\cite{1992CTVVpra,PQSCHA}, have found numerous applications in Physics and Physical Chemistry. Besides the tests on simple models with one degree of freedom~\cite{FeynmanKleinert1986,JankeK1986,JankeK1987,1990VTijmpb}, it is noteworthy that the very first paper regarded the 1D sine-Gordon model~\cite{GiachettiTognetti1985,1988GTVpra1}, whose classical version is characterized by the existence of topological nonlinear excitations, the solitons, that determine an anomaly of thermodynamic quantities like the specific heat: the GTFK method allowed for the first time to quantify the same anomaly for the quantum system, and was shown to agree with the outcomes of hard Quantum Monte Carlo calculations~\cite{1988GTVpra1} and to admit a renormalized continuum limit in agreement with exact `Bethe Ansatz' calculations~\cite{1988GTVpra3}.

Among many accomplishments, one should mention the quantitative explanation~\cite{1991CTVVprb} of experimental data regarding a quasi-1D magnet CsNiF$_3$, that behaves similarly to the sine-Gordon model, while a major one has been the study of 2D quantum anisotropic magnets~\cite{1995CTVVpra,1998CCTVVphd}, whose classical counterpart shows the topological phase transition studied~\cite{KosterlitzT1973} by Kosterlitz and Thouless (KT);
the GTFK approach allowed also to quantitatively characterize~\cite{2006CGVVjap} earlier experiments, showing that magnetic and calorimetric measurements performed in 1983 were the first known experimental observation of KT behavior in a real magnet; a further success in the magnetic realm was providing a consistent picture of the elusive Ising phase transition in a frustrated model such as the 2D quantum $J_1$-$J_2$ Heisenberg antiferromagnet~\cite{CapriottiFRT2004}.

2D Josephson-junction arrays are also typical KT systems: the effective potential was extended to include the dissipative effect of resistive shunts among the junctions used in experiments, getting quantitative accuracy for the phase diagram~\cite{2000CFTVprb}. The versatility of the GTFK potential is witnessed also by recent applications in the theoretical interpretation of thermal expansion measurements obtained by $x$-ray absorption spectroscopy in alloys~\cite{YokoyamaE2013,YokoyamaKU2018}.


\section*{Path-Integral formulation of Stochastic Calculus}

In this section, we briefly review how the formalism of stochastic calculus can be recast in the language of path-integrals in Euclidean time, focussing for simplicity on the case of a single SDE as in Eq.~(\ref{eq.diffusion}). As a first step, in order to simplify the derivation,  
it is convenient to transform the original process into an auxiliary one, $X_t$, with constant volatility $\sigma$. Following \cite{sahalia1999}, this 
can be achieved in general through the so-called Lamperti's transform
\begin{equation}
X_t = \gamma(Y_t) \equiv  \sigma \int^{Y_t}_{0}  \frac{dz}{\sigma_y(z)}~.
\label{inttransf}
\end{equation} 
A straightforward application of Ito's Lemma gives the stochastic differential equation satisfied by $X_t$ for $t\ge 0$:
\begin{equation}\label{eq.stproc}
d X_t = \mu(X_t) dt + \sigma d W_t, 
\end{equation} 
where
\begin{equation}
	\mu(x) = \sigma \left[\, \frac{\mu_y(\gamma^{-1}(x))}{\sigma_y(\gamma^{-1}(x))} - \frac{1}{2}\frac{\partial \sigma_y}{\partial y}(\gamma^{-1}(x))\right].
\label{driftX}
\end{equation}
Here, $y = \gamma^{-1}(x)$ is the inverse of the transformation (\ref{inttransf}).
The generalized AD density (\ref{eq.ad}) for the processes $X_t$ and $Y_t$ are related by the Jacobian associated with (\ref{inttransf}) giving 
\begin{align}
\psi_\lambda^Y(y_T, y_{0}, T) = \sigma \frac{\psi_\lambda(\gamma(y_T),x_0, T)}{\sigma_y (y_T)}~.
\end{align}

It is well known, see {\em e.g.}, \cite{andersen2010interest,karatzas1991brownian}, that the generalized AD density (\ref{eq.ad}) for the process (\ref{eq.stproc}) satisfies the following conjugate forward (Fokker-Planck)
partial differential equation (PDE)
\begin{align}\label{eq.fp}
\partial_t \psi_\lambda(x_t, x_0, t) = \Big( &-\lambda r(x) - \partial_x \mu(x_t) \nonumber \\ &+ \frac{1}{2}  \sigma^2\partial_x^2 \Big)   \psi_\lambda(x_t, x_0, t)~, 
\end{align}
with the initial condition $\psi_\lambda(x_t, x_0, 0) = \delta(x_0 - x_t)$.

\begin{figure}[t]
\centerline{\includegraphics[width=80mm]{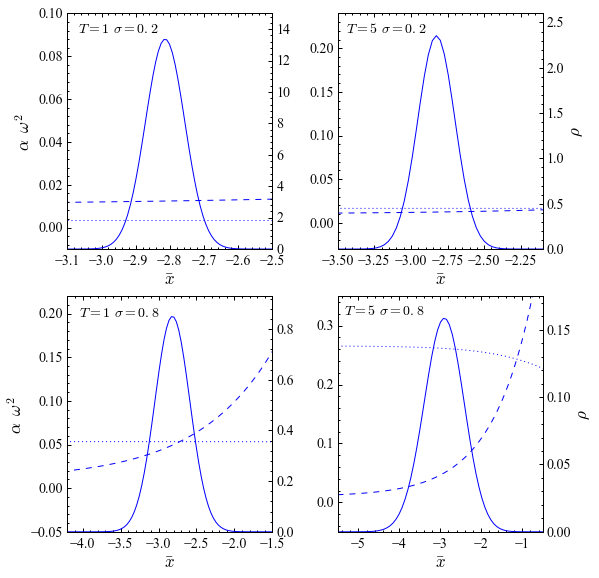}}
\vspace*{8pt}
\caption{Black-Karasinki model: GTFK self-consistent parameters (left axis) $\omega^2(\bar x)$ (dashed line), $\alpha(\bar x)$ (dotted line) and diagonal trial reduced density matrix $\bar \rho_{\bar x}(x_0, x_0, T)$ (right axis) as a function of the average point $\bar x$ for different values of the the time to maturity and volatility ({\em e.g.}, of the strength of the diffusive effects). The other parameters of the diffusion are mean-reversion level $a = 0.1$, speed $b=\ln 0.04$, and $x_0 = \ln 0.06$.}
\label{BKalphaomegarho}
\end{figure}

A path-integral representation of the AD density can be constructed  \cite{BennatiRosaClot1999} starting from the Euler approximation, correct up to $O(\Delta t)$, for the solution of
the Fokker-Planck PDE (\ref{eq.fp})
\begin{align}\label{eq.Euler}
&\psi_\lambda(x_{\Delta t},x_0,\Delta t) = e^{-\lambda r\left(x_0\right)\Delta t} \times \nonumber \\ 
&\frac{1}{\sqrt{2\pi\sigma^2\Delta t}} \exp\left[ -\frac{(x_{\Delta t}-x_0-\mu\left(x_0\right)\Delta t)^2}{2\sigma^2\Delta t}\right]~.
\end{align}
Using the Markov property, the equation above gives a prescription to write the solution of the Fokker-Planck equation in the form of a convolution product of short-time
AD densities as:
\begin{align}\label{eq.discrete}
&\psi_\lambda(x_T,x_0,T) = \left(\frac{1}{{2\pi\sigma^2\Delta t}}\right)^{N/2}  \times \nonumber \\ &\int \ldots \int \prod_{i=1}^{N-1}dx_i  \exp{\big[\tilde S(x_i,x_{i-1})\big]}~,
\end{align}
with $\Delta t = T / N$, $x_N\equiv x_T$  and 
\begin{align}\label{eq.discaction}
&\tilde S(x_i,x_{i-1}) = -\frac{\Delta t}{2\sigma^2}\left [ \frac{(x_i - x_{i-1})}{\Delta t} + \mu((x_{i-1}+x_i)/2)\right]^2  - \nonumber \\
& \Delta t \big[\partial_x \mu((x_{i-1}+x_{i})/2)/2 +\lambda r((x_{i-1}+x_{i})/2)\big]~,
\end{align}
where the term 
\begin{equation}
\Delta t \partial_x \mu((x_{i-1}+x_{i})/2)/2~, 
\end{equation}
arises, at order $O(\Delta t)$, from using the analytically convenient Stratonovich mid-point discretization \cite{BennatiRosaClot1999}.
As a result, the limit $N\to \infty$ of Eq.~(\ref{eq.discrete}) can be formally written as
\begin{equation}\label{eq.adpi}
\psi_\lambda(x_T,  x_0, T)  =  e^{-W(x_T, x_0)}  \rho(x_T,  x_0, T)~,
\end{equation}
where 
\begin{equation}\label{eq.piSC}
\rho(x_T,  x_0, T) = \int_{x(0) =x_0}^{x(T) = x} \hspace{-0.1cm}{\cal D} [x(t)] \, \,e^{S[x(t)]}~,
\end{equation}
has the same form of the {\em density matrix} in  Eq.~(\ref{eq.pi}), the functional
\begin{equation}\label{eq.actionSC}
S[x(t)] = - \int_{0}^T dt \left[ \frac{1}{ 2\sigma^2 } \dot x^2(t) + V(x(t)) \right]~,
\end{equation}
has the same form of the {\em euclidean action} in Eq.(\ref{eq.action}), 
\begin{equation}\label{eq.driftpot}
V(x) = \frac{\mu(x)^2}{2\sigma^2} + \frac{\mu^\prime(x)}{2} + \lambda r(x)~,
\end{equation}
can be called {\em drift potential}  and we have defined
\begin{equation}
W(x_T, x_0) = - \frac{1}{\sigma^2} \int_{x_0}^{x_T}  dx \,\,\mu(x)~,
\end{equation}
in order to give Eq.~(\ref{eq.actionSC}) a suggestive Lagrangian structure as in Eq.~(\ref{eq.action}). 

The key observation is that the path integral in Eq.~(\ref{eq.piSC}) is formally equivalent to density matrix in Eq.~(\ref{eq.pi}) describing the quantum termodynamics of a particle of mass $m = \hbar/\sigma^2$ in a potential $\hbar V(x)$, at temperature  ${\cal T}= \hbar / k_B T$ (such that $\beta \hbar = T$). 

The GTFK can be therefore applied straightforwardly and here for convenience we restate the results with the notation of stochastic calculus:
\begin{align}\label{eq.reddens2SC}
& \bar \rho_{\bar x}(x_T, x_0, T)  =  \sqrt{\frac{1}{2\pi\sigma^2T}} e^{-Tw(\bar x)} \frac{f}{\sinh f} \times  \nonumber \\ 
&\frac{1}{\sqrt{2\pi\alpha}} \exp\left[ -\frac{\xi^2}{2\alpha} -\frac{\omega}{4\sigma^2}\coth f (x_T-x_0)^2 \right]~,
\end{align}
where $\xi = (x_T+x_0)/2 - \bar x$, $f = \omega(\bar x) T/2$ and 
\begin{equation}\label{eq.alphaSC}
\alpha(\bar x) = \frac{\sigma^2}{2\omega(\bar x)}\left (\coth f(\bar x) -\frac{1}{f(\bar x)} \right )~,
\end{equation}
with $w(\bar x)$ and $\omega(\bar x)$ solutions of the self-consistent equations:
\begin{align} 
\langle \langle V(\bar x + \xi) \rangle\rangle &=  \langle \langle V_{\bar x}(\bar x + \xi) \rangle\rangle \nonumber \\ & = w(\bar x) + \frac{\omega^2(\bar x)\alpha(\bar x)}{2\sigma^2}~, \label{GTFK1_SC}\\ 
\langle \langle V^{\prime\prime}(\bar x + \xi) \rangle\rangle&=\langle \langle  V_{\bar x}^{\prime\prime}(\bar x + \xi)\rangle\rangle = \frac{\omega^2(\bar x)}{\sigma^2}~.\label{GTFK2_SC}
\end{align}

The GTFK method, becomes exact in the limit of short time to maturity $T\to 0$ and vanishing volatility $\sigma \to 0$ for which the parameter $\alpha$ vanishes as $\sigma^2 T/12$.
Furthermore, given the form of the chosen trial potential, for harmonic actions, the GTFK approximation is, in fact, exact. This is for instance the case for the Vasicek model \cite{Vasicek1977} as it will be illustrated in the next section.

\begin{figure}[t]
\centerline{\includegraphics[width=80mm]{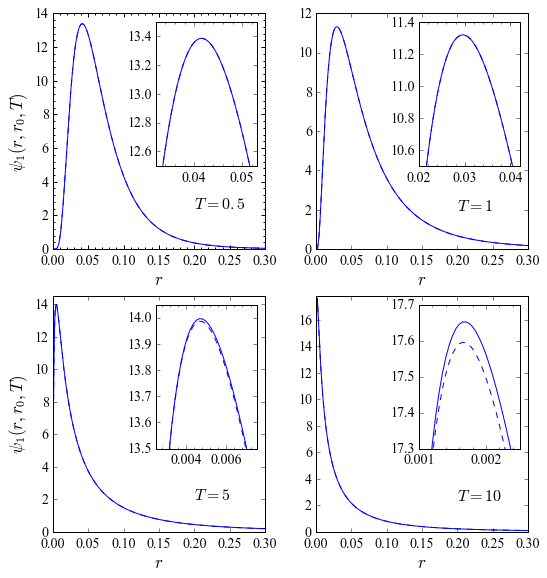}}
\vspace*{8pt}
\caption{Black-Karasinki AD densities obtained with the GTFK method (dashed line) and a numerical solution of the Fokker-Plank PDE  (continuos line) for different values of the the time to maturity. 
The parameters of the BK process are: mean-reversion speed  $a = 0.1$, level $b = \ln 0.04$, volatility $\sigma = 0.85$, and initial rate $r_{0} = 0.060$. The inset is an enlargement of the region of the maximum where the discrepancy between the PDE result and GTFK approximation is largest.}
\label{BKpsi}
\end{figure}

\section*{Numerical Results}

In this section we illustrate the effectiveness of the GFTK approach by discussing its application to a few diffusions processes of the form (\ref{eq.diffusion}), starting from two cases in which the method gives
exact results, namely the Vasicek and the so-called quadratic short-rate model. We then discuss the Black-Karasinski (BK) \cite{BK} and GARCH linear SDE model \cite{MercurioGarch, EEGARCH} - for which the AD density (\ref{eq.ad}) or zero-coupon bonds (\ref{eq.zeroad}) are not know analytically - by presenting the comparison of the GTFK results with those obtained by solving numerically the relevant PDEs and  by employing other approximations.

\subsection{Vasicek model}
The Vasicek model \cite{Vasicek1977} is a simple example of affine process \cite{duffie}
\begin{equation}\label{eq.OU}
dX_t = a(b-X_t) dt + \sigma dW_t
\end{equation}
where $a$ is the mean-reversion speed, $b$ the mean-reversion level, $\sigma$ the volatility, and $r(X_t) = X_t$. The drift potential (\ref{eq.driftpot}) is given by the quadratic form
\begin{equation}
V_V(x) = \frac{a^2 (b-x)^2}{2\sigma^2} - \frac{a}{2} + \lambda x~.
\end{equation}
The path integral for quadratic potentials is known to be analytically tractable and corresponds in quantum Physics 
to the so-called {\em harmonic oscillator} \cite{feynman2010quantum}. In this case, the GTFK self-consistent 
conditions (\ref{GTFK1_SC}) and (\ref{GTFK2_SC}) read:
\begin{align}
w(\bar x) = V_V(\bar x)~,\,\,\,\,\,\,\,  \omega^2(\bar x) =  a^2~,
\end{align}
and the reduced density matrix (\ref{eq.reddens2SC}) reads:
\begin{align}\label{eq.reddensvasicek}
& \bar \rho_{\bar x}(x_T, x_0, T)  =  \sqrt{\frac{1}{2\pi\sigma^2T}} e^{-T V_V(\bar x)} \frac{f}{\sinh f} \times  \nonumber \\ 
&\frac{1}{\sqrt{2\pi\alpha}} \exp\left[ -\frac{\xi^2}{2\alpha} -\frac{a}{4\sigma^2}\coth f (x_T-x_0)^2 \right]~,
\end{align}
with $\alpha = \sigma^2/2 a (\coth f - 1/f)$, $f = aT/2$, both independent of $\bar x$. The integral over $\bar x$ in Eq.~(\ref{eq.PIADprice}) can then be 
performed analytically giving, after a somewhat tedious but straightforward calculation,
\begin{align}\
&\psi_\lambda(x_T, x_0, T) = \frac{1}{\sqrt{2\pi\bar \sigma^2}} e^{{\lambda}(x-x_0)/a} e^{-T(\lambda b - \lambda^2 \sigma^2 / 2 a^2) } \times \nonumber \\ &
\exp\left[ -\frac{\big((x_T - b +  \frac{\lambda\sigma^2}{a^2}) - (x_0 - b +  \frac{\lambda\sigma^2}{a^2}) e^{-aT}\big)^2}{2\bar\sigma^2}   \right]
\end{align}
where $\bar \sigma^2 = \sigma^2 (1 - \exp(-2 a T))/ 2 a$,  in agreement with the known result \cite{Jashmidian1989}. 

\subsection{Quadratic Short Rate Model}

In the quadratic short rate model, the short rate is defined as 
\begin{equation}\label{eq.qr}
r(X_t) =  1 + \beta X_t + \gamma X_t^2~,
\end{equation}
with $X_t$ following the OU diffusion (\ref{eq.OU}), which is positive definite for $\beta>0$ and $\gamma^2<4\beta$. In this case, the drift potential (\ref{eq.driftpot}) reads
\begin{equation}
V_Q(x) = \frac{a^2(b-x)^2}{2\sigma^2} -\frac{a}{2} + \lambda (1 + \beta x + \gamma x^2)~,
\end{equation}
while the GTFK conditions,  (\ref{GTFK1_SC}) and (\ref{GTFK2_SC}), can be determined as
\begin{equation}
w(\bar x) = V_Q(\bar x)~,\,\,\,\,\,\,\,  \omega^2(\bar x) =  a^2 + 2 \lambda \gamma  \sigma^2~,
\end{equation}
which, as in the Vasicek model discussed above give a frequency $\omega$ that is not dependent on the average point and a function $w(\bar x)$ which is 
quadratic in $\bar x$.  Also in this case the Gaussian integration can be performed analytically leading to the exact result.

\subsection{Black-Karasinki Model}

The BK \cite{BK} model is a conspicuous example of a diffusion that is particularly suitable 
for financial applications because the short rate at any time horizon follows an intuitive lognormal distribution.  
Unfortunately,  it lacks the same degree of analytical tractability as that shown by affine models. 
As a result, although widely used in practice, BK implementations rely on computationally intensive  numerical simulations based on PDE or
Monte Carlo \cite{andersen2010interest}.

The short rate in the BK model is defined as 
 \begin{equation}
r(X_t) =  \exp{X_t}~,
\end{equation}
with $X_t$ following the OU diffusion (\ref{eq.OU}). In this case, the drift potential (\ref{eq.driftpot}) reads
\begin{equation}
V_{BK}(x) = \frac{a^2(b-x)^2}{2\sigma^2} -\frac{a}{2} + \lambda e^x~,
\end{equation}
while the GTFK conditions,  (\ref{GTFK1_SC}) and (\ref{GTFK2_SC}), can be determined with some straightforward algebra as
\begin{align}
w(\bar x) &= V_{BK}(\bar x)  + \frac{a^2-\omega^2(\bar x)}{\sigma^2}\alpha(\bar x)  \nonumber \\ & + \lambda \left(e^{\alpha(\bar x)/2} - 1\right) e^{\bar x} ~, \\
\omega^2(\bar x) &= a^2 + \lambda \sigma^2e^{\alpha(\bar x)/2}e^{\bar x}~,
\end{align}
with the second to be solved self-consistently with the renormalization parameter in Eq.~(\ref{eq.alphaSC}).

In Fig.~\ref{BKalphaomegarho} we plot the GTFK self-consistent parameters  $\omega^2(\bar x)$, and $\alpha(\bar x)$ and the diagonal trial reduced density matrix $\bar \rho_{\bar x}(x_0, x_0, T)$ 
in Eq.~(\ref{eq.reddens2SC}) as a function of the average point $\bar x$ for different strength of the diffusive effects, namely of  the time to maturity and volatility. For weak diffusive effects, the parameter $\alpha(\bar x)$ is relative small and the trial reduced density matrix has a sharp peak around $x_0$. In this region, both $\alpha(\bar x)$ and $\omega^2(\bar x)$ display a weak dependence on $\bar x$ which signals the adequacy of a local harmonic approximation to capture the purely diffusive effects in the problem. However, as the diffusive effects increase, with larger volatility and/or time to maturity, the renormalization parameter $\alpha(\bar x)$ increases, the trial density broadens and both  $\alpha(\bar x)$  and $\omega^2(\bar x)$ display a more marked dependency on the average point $\bar x$, signaling that a non-local approximation is needed to best capture the  diffusive effects given an harmonic ansatz of the effective potential.

An illustration of the accuracy of the BK AD densities (\ref{eq.ad}) obtained with the GTFK approximation is displayed for a high volatility case  in Fig.~\ref{BKpsi}, for different values of time to maturity, by comparing with a numerical solution of the Fokker-Planck equation (\ref{eq.fp}). Here we observe that the GTFK approximation is hardly distinguishable from the PDE result up to $T=5$, and remains 
very accurate even for large time horizons. 

\begin{figure}[t]
\centerline{\includegraphics[width=80mm]{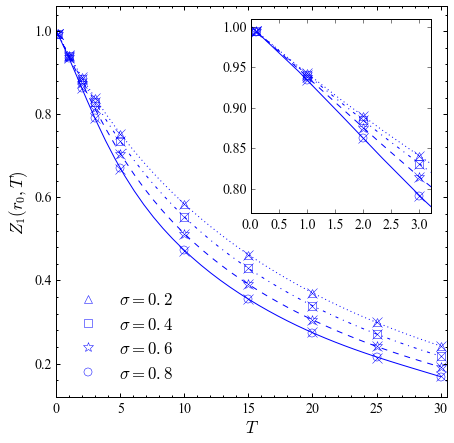}}
\vspace*{8pt}
\caption{GTFK zero-coupon bond prices as a function of time to maturity for the Black-Karasinski model, with mean-reversion speed $a = 0.1$, level $b = \ln 0.04$, initial rate $r_{0} = 0.06$, and different values of the volatility. Crosses indicate the PDE results. The inset is an enlargement for short times to maturity.}
\label{zeroiBK}
\end{figure}

This is also confirmed by the results for zero-coupon bonds (\ref{eq.zeroad}) reported in Table \ref{tablevsEE} illustrating how the GTFK method compares favorably with the results obtained with recently proposed semi-analytical approximations, namely the Exponent Expansion (EE) \cite{EEBK}, and the  Karhunen-Lo\'eve (KL) expansions \cite{daniluk2016} when benchmarked agains a numerical solution of the associated PDE. In particular, for short time horizons, the GTFK approximation  has comparable accuracy  with the  EE. For larger time horizons, the GTFK compares better and better and remains very accurate  even when the EE, which has a finite convergence ratio in $T$, eventually breaks down. Similarly, the GTFK method has better accuracy than the first order KL expansion, and comparable accuracy with the second order KL expansion for short time horizons, while it has significantly better accuracy for large time horizons.  Even for time horizons as large as 20 years the GTFK approximation produces zero-coupon bond prices within 50 basis points from the exact result, as also illustrated in Fig.~\ref{zeroiBK}. Similar conclusions can also be drawn when comparing with other recently proposed approaches as those in Refs.~\cite{Hagan07, AntonovSpector2011}.

\begin{widetext}

\begin{table}[t]
{\begin{tabular}{@{}lccccc@{}}  \toprule
 $T$  &    ${\rm EE}$      & KL(1)                   &   KL(2)                 &   GTFK      &    PDE         \\ \colrule
 0.1    &   0.9939  (0.00\%)  &  0.9939 (0.00\%) &  0.9939 (0.00\%) &  0.9939 (0.00\%) &    0.9939     \\  
 0.5    &   0.9681  (0.00\%)  &  0.9681 (0.00\%) &  0.9681 (0.00\%) &  0.9681 (0.00\%) &    0.9681     \\  
 1.0    &   0.9331  (0.00\%)  &  0.9331 (0.00\%) &  0.9331 (0.00\%)  &  0.9331 (0.00\%) &    0.9331     \\
 2.0    &   0.8581  (0.01\%)  &  0.8580 (0.02\%) &  0.8581 (0.01\%)  &  0.8582 (0.00\%) &    0.8582     \\
 3.0    &   0.7845  (0.01\%)  &  0.7842 (0.05\%) &  0.7844 (0.02\%)  &  0.7847 (0.01\%) &    0.7846     \\
 5.0    &   0.6595  (0.04\%)  &  0.6582 (0.24\%) &  0.6593 (0.08\%)  &  0.6602 (0.06\%) &    0.6598     \\ 
 10.0  &           -                   &  0.4545 (1.69\%)  & 0.4601 (0.48\%)  &   0.4628 (0.10\%) &    0.4623      \\
 20.0  &           -                   &  0.2440 (9.06\%)  & 0.2592 (3.38\%)  &   0.2672 (0.41\%) &    0.2683      \\
\botrule
\end{tabular} }
\caption{Black-Karasinski  $T$ maturity zero-coupon bonds obtained with the GTFK approximation, the Exponent Expansion (EE) of Ref.~\cite{EEBK}, the Karhunen-Lo\'eve (KL) expansion of Ref.~\cite{daniluk2016} to first and second order, and by solving numerically the associated PDE. The parameters of the BK process are: mean-reversion speed  $a = 0.1$, level $b = \ln 0.04$, volatility $\sigma = 0.85$, and initial rate $r_{0} = 0.06$.}
\label{tablevsEE} 
\end{table}  

\end{widetext}

\subsection{GARCH Linear SDE}

As an example of a more challenging application, we then consider the GARCH linear SDE or Inhomogenous Geometric Brownian Motion \cite{MercurioGarch,EEGARCH} model, which is a special case of 
the so-called Continuous Elasticity of Variance (CEV) diffusion \cite{CEV}, namely
\begin{equation}\label{eq.garch}
dY_t = a(b-Y_t)dt + \sigma Y_t dW_t~,
\end{equation}
with $r(Y_t) =  Y_t$.

The process defined by the SDE in Eq.~(\ref{eq.garch}) can be shown to be strictly positive \cite{KloedenPlaten}. As a result, like the BK model, it is well suited to represent default intensities. It can be also shown to have  probability density profiles which are more intuitive than those generated by the widely used square-root processes \cite{cir,MercurioGarch}. Unfortunately, even if it can be solved exactly \cite{KloedenPlaten} it does not admit a closed form for the (generalized) AD prices (\ref{eq.ad}). 

\begin{figure}[t]
\centerline{\includegraphics[width=80mm]{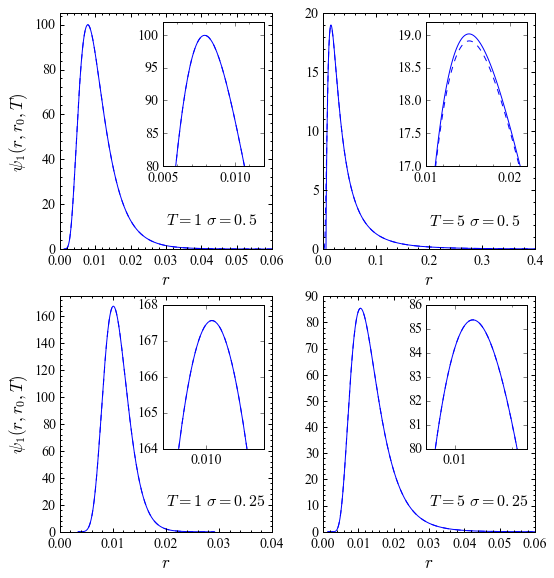}}
\vspace*{8pt}
\caption{GARCH linear SDE AD densities obtained with the GTFK method (dashed line) and a numerical solution of the Fokker-Plank PDE  (continuos line) for different values of the the time to maturity and volatility.  The other parameters of the process are: mean-reversion speed  $a = 0.1$, level $b = 0.02$, and initial rate $y_{0} = 0.01$. The inset is an enlargement of the region of the maximum where the discrepancy between the PDE result and GTFK approximation  is the largest.}
\label{GARCHpsi}
\end{figure}

Under the Lamperti's transformation (\ref{inttransf}) for this process, namely $X_t = \log Y_t$, Eq.~(\ref{eq.garch}) reads
\begin{equation}\label{eq.garchLam}
dX_t = \mu_G(X_t) dt +\sigma dW_t~,  
\end{equation}
with 
\begin{equation}
\mu_G(x) = a b \,e^{-x} - a -\sigma^2/2~.
\end{equation}
The drift potential (\ref{eq.driftpot}) associated with the SDE (\ref{eq.garchLam}) reads therefore
\begin{align}\label{eq.Gpot}
V_{G}(x) &=  \frac{a^2b^2}{2\sigma^2} e^{-2x} - \frac{ab}{\sigma^2}e^{-x}(a+\sigma^2)+ \nonumber\\
&\frac{1}{2\sigma^2}(a^2+{\sigma^2}/{2})^2  + \lambda e^{x}~,
\end{align}
which is related to the so-called Morse potential \cite{morse}. The GTFK conditions,  (\ref{GTFK1_SC}) and (\ref{GTFK2_SC}), can be determined 
with some straightforward algebra as
\begin{align}
&w(\bar x) = \frac{a^2b^2}{2\sigma^2} e^{-2x} e^{2\alpha} - \frac{ab}{\sigma^2}e^{-x}(a+\sigma^2) e^{\alpha/2}+ \nonumber  \\
&\frac{1}{2\sigma^2}(a^2+{\sigma^2}/{2})^2 + \lambda e^{x}e^{\alpha/2}-\frac{\omega^2(\bar x)\alpha(\bar x)}{2\sigma^2}~ \\
&\omega^2(\bar x) = 2 a^2 b^2 e^{-2x} e^{2\alpha}  - ab e^{-x}(a+\sigma^2) e^{\alpha/2} + \nonumber \\& \lambda \, \sigma^2  e^{x}e^{\alpha/2}~.
\end{align}

Examples of AD densities (\ref{eq.ad}) obtained with the GTFK approximation for the GARCH linear SDE are displayed in Fig.~\ref{GARCHpsi}, for different values of the diffusion parameters, with a comparison
with a  numerical solution of the Fokker-Planck equation (\ref{eq.fp}). Here we observe that the GTFK approximation, as in the BK case, is difficult to distinguish from the PDE result up to several years maturity, and for large enough volatilities. As in the BK case, the accuracy of the approximations depends on the chosen model parameters, and the maturity being considered. The approximation becomes less accurate for larger maturities $T$ and volatility. The behaviour with respect to the mean-reversion speed $a$ is instead less clear-cut as this parameter affects both the variance of the process and the non-linearity of the drift potential
(\ref{eq.Gpot}).

The accuracy of the GTFK method for the GARCH linear SDE is also illustrated for zero-coupon bonds (\ref{eq.zeroad}) in Table \ref{tableGARCH} and \ref{tableGARCH2} for two sets of model parameters, showing how the GTFK method compares favorably with the results obtained with recently proposed semi-analytical approximations, namely the EE \cite{EEGARCH}, 
when benchmarked agains a numerical solution of the associated PDE. In general, although less accurate than in the BK case, due to the more complex form of the drift potential (\ref{eq.Gpot}),  the approximation produces satisfactory results for maturities up to several years even in regimes of high volatility.

\begin{table}[t]
{\begin{tabular}{@{}lccc@{}}  \toprule
 $T$  &    ${\rm EE}$      &  GTFK      &    PDE         \\ \colrule
 0.1    &   0.9940 (0.00\%) &    0.9940 (0.00\%) &    0.9940     \\  
 0.5    &   0.9707 (0.00\%)  &   0.9707 (0.00\%) &    0.9707     \\  
 1.0    &   0.9429 (0.00\%)  &   0.9429 (0.00\%) &    0.9429     \\
 2.0    &   0.8914 (0.03\%)  &   0.8920 (0.03\%) &    0.8917     \\
 3.0    &   0.8459 (0.08\%)  &   0.8472 (0.07\%) &    0.8466     \\
 5.0    &   0.7834 (1.40\%)  &   0.7717 (0.12\%) &    0.7726     \\ 
 7.5    &   -                          &   0.6923 (1.45\%) &    0.7025     \\
 10.0  &   -                          &   0.6223 (3.92\%) &    0.6477     \\
\botrule
\end{tabular} }
\caption{GARCH linear SDE $T$ maturity zero-coupon bonds obtained with the GTFK approximation, the Exponent Expansion (EE) of Ref.~\cite{EEGARCH}, and by solving numerically the associated PDE. The parameters of the process are: mean-reversion level $a = 0.1$, level $b = 0.04$, volatility $\sigma = 0.6$, and initial rate $y_0 = 0.06$.}
\label{tableGARCH} 
\end{table}  

\begin{table}[t]
{\begin{tabular}{@{}lccc@{}}  \toprule
 $T$  &    ${\rm EE}$      &  GTFK      &    PDE         \\ \colrule
 0.1     &    0.9990 (0.00\%)  &    0.9990 (0.00\%) &  0.9990       \\  
 0.25   &    0.9975 (0.00\%)  &    0.9975 (0.00\%) &  0.9975       \\  
 0.5     &    0.9949 (0.00\%)  &    0.9949 (0.00\%) &  0.9949      \\  
 1.0     &    0.9896 (0.00\%)  &    0.9896 (0.00\%) &  0.9896       \\   
 2.5     &    0.9728 (0.02\%)  &    0.9723 (0.03\%) &  0.9726       \\
 5.0     &    0.9359 (0.62\%)  &    0.9403 (0.15\%) &  0.9417      \\ 
 10.0   &    0.8315 (5.10\%)  &    0.8709 (0.60\%)   &  0.8762       \\
\botrule
\end{tabular} }
\caption{GARCH linear SDE  $T$ maturity zero-coupon bonds obtained with the GTFK approximation, the Exponent Expansion (EE) of Ref.~\cite{GaukharThesis}, and by solving numerically the associated PDE. The parameters of the process are: mean-reversion level $a = 0.1$, level $b = 0.02$, volatility $\sigma = 0.5$, and initial rate $y_0 = 0.01$.}
\label{tableGARCH2} 
\end{table}

\section*{Conclusions}

An effective-potential path-integral formalism of quantum statistical mechanics -- dubbed GTFK after the authors \cite{GiachettiTognetti1985, FeynmanKleinert1986} who  originally introduced it -- has been widely utilized in Physics for the study of the quantum thermodynamics of condensed matter systems. The method is based on a self-consistent harmonic approximation of the pure-quantum contributions to the thermodynamics, while fully accounting for the classical behaviour of the system \cite{PQSCHA}. As a semiclassical approach, it is exact in the high-temperature and zero-quantum fluctuations limits  but, remarkably, it also gives a meaningful representation  in the zero-temperature limit, where it is equivalent to a self-consistent harmonic approximation of the potential.

By exploiting the path-integral formulation of stochastic calculus, we have shown how the GTFK approach can be used to develop an accurate semi-analytical  approximation of (generalized) Arrow-Debreu densities, and zero-coupon bonds for non-linear diffusions.  The method is exact in the limit of zero volatility, zero time to maturity, and for Ornstein-Ulhenbeck diffusions.

The GTFK provides remarkably accurate results for the Black-Karasinski and GARCH linear SDE for interest rates or default intensities, even for high volatilities and long time horizons, with results that compare favorably with previously presented approximation schemes \cite{Hagan07, AntonovSpector2011, daniluk2016, EEBK, EEGARCH}, with expressions that are more compact and easier to 
compute, and less severe limitations arising from a finite convergence radius in the time to maturity or volatility. Similarly to the approach in \cite{Capriotti2006}, the range of application of the expansion can be further extended to even larger time horizons by means of a fast numerical convolution \cite{BennatiRosaClot1999}. 

The GTFK approximation can be potentially improved in one of two ways: by pursuing higher-order corrections as in the so-called variational perturbation theory \cite{kleinert2009path} or by its generalization
to Hamiltonian systems \cite{1992CTVVpra,PQSCHA} that would allow avoiding the non-linearities in the potential introduced ({\em e.g.}, as for the GARCH linear SDE) via the Lamperti's transformation (\ref{inttransf}). 

The accuracy and ease of computation of the GTFK method makes it a computationally efficient alternative to fully numerical schemes such as binomial trees, PDE or Monte Carlo for the 
calculation of transition densities -- whether for the maximization of  classical likelihoods or the computation of posterior distributions --  and for the evaluation of European-style derivatives. 
This is of practical utility {\em e.g.}, for econometric applications \cite{sahalia1999}, for speeding up pricing or calibration routines for valuation of derivatives \cite{andersen2010interest} or in the context of time consuming multi-factor simulations that are common  place in  financial engineering in a variety of applications \cite{hull2017options}.

\begin{acknowledgements}

It is a pleasure to acknowledge Jim Gatheral, Tao-Ho Wang and Mehdi Sonthonnax for useful discussions.  
The authors are grateful to Prof. Valerio Tognetti for igniting in them the passion for Path Integrals, and for his warm 
support throughout the years.

\end{acknowledgements}

\appendix

\bibliography{biblio}     

\end{document}